\documentclass[a4paper,
                keeplastbox,   
              ]{jacow}
\makeatletter%
	\ifboolexpr{bool{xetex}}
	 {\renewcommand{\Gin@extensions}{.pdf,%
	                    .png,.jpg,.bmp,.pict,.tif,.psd,.mac,.sga,.tga,.gif,%
	                    .eps,.ps,%
	                    }}{}
\makeatother

\ifboolexpr{bool{xetex} or bool{luatex}} 
 {}                                      
 {\usepackage[utf8]{inputenc}}           

\usepackage[USenglish]{babel}			

\ifboolexpr{bool{jacowbiblatex}}%
 {%
  \addbibresource{jacow-test.bib}
  \addbibresource{biblatex-examples.bib}
 }{}
\usepackage[caption=false]{subfig}
	\DeclareCaptionListOfFormat{myparens}{#2}
\newcommand{\Subref}[1]{\protect\subref{#1}}
\usepackage{xspace}
\usepackage{color}
\usepackage{xcolor}
  \definecolor{white}{rgb}{1,1,1}
  \definecolor{red}{rgb}{1,0,0}
  \definecolor{blue}{rgb}{0,0,1}
  \definecolor{darkBlue}{rgb}{0,0,.4}
  \definecolor{green}{rgb}{0,1,0}
  \definecolor{magenta}{rgb}{1,0,.6}
  \definecolor{lightblue}{rgb}{0,.5,1}
  \definecolor{lightpurple}{rgb}{.6,.4,1}
  \definecolor{gold}{rgb}{.6,.5,0}
  \definecolor{orange}{rgb}{1,0.4,0}
  \definecolor{hotpink}{rgb}{1,0,0.5}
  \definecolor{newcolor2}{rgb}{.5,.3,.5}
  \definecolor{newcolor}{rgb}{0,.3,1}
  \definecolor{newcolor3}{rgb}{1,0,.35}
  \definecolor{darkgreen1}{rgb}{0, .35, 0}
  \definecolor{darkgreen}{rgb}{0, .6, 0}
  \definecolor{darkred}{rgb}{.75,0,0}
  \xdefinecolor{olive}{cmyk}{0.64,0,0.95,0.4}
  \xdefinecolor{purpleish}{cmyk}{0.75,0.75,0,0}
  \definecolor{grayBold}{rgb}{0,0,0}
  \definecolor{battleshipgrey}{rgb}{0.52, 0.52, 0.51}
  \definecolor{cinereous}{rgb}{0.6, 0.51, 0.48}
  \definecolor{darkgray}{rgb}{0.66, 0.66, 0.66}
  \definecolor{davysgrey}{rgb}{0.33, 0.33, 0.33}
  \definecolor{dimgray}{rgb}{0.41, 0.41, 0.41}
  \definecolor{gainsboro}{rgb}{0.86, 0.86, 0.86}
  \definecolor{grullo}{rgb}{0.66, 0.6, 0.53}
  \definecolor{manatee}{rgb}{0.59, 0.6, 0.67}
  \definecolor{oldlavender}{rgb}{0.47, 0.41, 0.47}
  \definecolor{oldmauve}{rgb}{0.4, 0.19, 0.28}
  \definecolor{khaki}{rgb}{0.76, 0.69, 0.57}
  \definecolor{paynesgrey}{rgb}{0.25, 0.25, 0.28}
  \definecolor{platinum}{rgb}{0.9, 0.89, 0.89}
  \definecolor{whitesmoke}{rgb}{0.96, 0.96, 0.96}
  \definecolor{coolblack}{rgb}{0.0, 0.18, 0.39}
  \definecolor{royalblue}{rgb}{0.25, 0.41, 0.88}
  \definecolor{red0}{RGB}{243,61,99}
  \definecolor{red1}{RGB}{198,40,40}
  \definecolor{green0}{RGB}{139,195,74}
  \definecolor{blue0}{RGB}{33,150,243}
  \definecolor{blue1}{RGB}{57,73,171}
  \definecolor{blue2}{RGB}{134, 165, 216}

\newcommand{\FigRef}[1]{\ensuremath{{\txt{Figure~\ref{#1}}}}\xspace}
\newcommand{\figRef}[1]{\ensuremath{{\txt{Fig.~\ref{#1}}}}\xspace}
\newcommand{\FigSubRef}[2]{\ensuremath{{\txt{Figure~\ref{#1}\Subref{#2}}}}\xspace}
\newcommand{\figSubRef}[2]{\ensuremath{{\txt{Fig.~\ref{#1}\Subref{#2}}}}\xspace}

\newcommand{\figsSubRef}[2]{\ensuremath{{\txt{Figs.~\ref{#1}\Subref{#2}}}}\xspace}

\newcommand{\hlt}[1]{\textbf{#1}\xspace}

\newcommand{\under}{\kern-0.1em\_\kern-0.1em}

\newcommand{\Sim}{\sim\kern-0.2em\xspace}

\newcommand{\txt}[1]{\text{#1}}
\newcommand{\tit}[1]{\textit{#1}}
\newcommand{\ttt}[1]{\texttt{#1}}

\newcommand{\eg}{e.g.\/,\xspace}

\newcommand{\djs}{\texttt{d3.js}\xspace}

\newcommand{\jvs}{\texttt{Javascript}\xspace}

\newcommand{\pyramid}{\texttt{Pyramid}\xspace}
\newcommand{\webSock}{\texttt{Web Sockets}\xspace}
\newcommand{\polymer}{\texttt{Polymer}\xspace}

\newcommand{\webComp}{\texttt{Web Component}\xspace}
\newcommand{\google}{\texttt{Google}\xspace}
\newcommand{\redis}{\texttt{redis}\xspace}

\newcommand{\python}{\texttt{Python}\xspace}
\newcommand{\cpp}{\texttt{C++}\xspace}
\newcommand{\java}{\texttt{Java}\xspace}

\newcommand{\gev}{GeV\xspace}
\newcommand{\tev}{TeV\xspace}

\newcommand{\gui}{GUI\xspace}
\newcommand{\acs}{ACS\xspace}
\newcommand{\hci}{HCI\xspace}

\newcommand{\iact}{IACT\xspace}
\newcommand{\iacts}{IACTs\xspace}

\newcommand{\lst}{\txt{LST}\xspace}
\newcommand{\lsts}{\txt{LSTs}\xspace}
\newcommand{\mst}{\txt{MST}\xspace}
\newcommand{\msts}{\txt{MSTs}\xspace}
\newcommand{\sst}{\txt{SST}\xspace}
\newcommand{\ssts}{\txt{SSTs}\xspace}
\newcommand{\oes}{\txt{OES}\xspace}

\newcommand{\cta}{CTA\xspace}
\newcommand{\hess}{H.E.S.S.\xspace}
\newcommand{\magic}{MAGIC\xspace}
\newcommand{\veritas}{VERITAS\xspace}
\newcommand{\alma}{ALMA\xspace}

\newcommand{\too}{\txt{ToO}\xspace}

\newcommand{\gamray}{$\gamma$-ray\xspace}
\newcommand{\gamrays}{$\gamma$-rays\xspace}

\usepackage{geometry}
\usepackage{fancyhdr}
\usepackage{ifthen}
\newboolean{isDraft} \setboolean{isDraft}{false}

\ifthenelse{\boolean{isDraft}}{
  \usepackage[ddmmyyyy]{datetime}
  \cfoot{\textcolor{red0}{\bf{- page \thepage\;-  {\sffamily DRAFT} - \today~(\currenttime) -}}}                   
}{}

\begin{document}

\title{The Graphical User Interface of the Operator of the Cherenkov Telescope Array}


\author{
  I.~Sadeh\thanks{\protect\url{iftach.sadeh@desy.de}},
  I.~Oya, DESY-Zeuthen, D-15738 Zeuthen, Germany\\
  J.~Schwarz, INAF - Osservatorio Astronomico di Brera, Italy\\
  E.~Pietriga, INRIA Saclay - Ile de France, LRI (Univ. Paris-Sud \& CNRS), France\\
  D.~De\v{z}man, Cosylab d.\/d.\/, Gerbi\v{c}eva 64, 1000 Ljubljana, Slovenia\\
  for the CTA Consortium\thanks{\protect\url{http://www.cta-observatory.org/}}
}

	
\maketitle

\begin{abstract}
  The Cherenkov Telescope Array (\cta) is the next generation gamma-ray observatory. \cta will incorporate about 100 imaging atmospheric Cherenkov telescopes (\iacts) at a southern site, and about 20 in the north. Previous \iact experiments have used up to five telescopes. Subsequently, the design of a graphical user interface (\gui) for the operator of \cta poses an interesting challenge. In order to create an effective interface, the \cta team is collaborating with experts from the field of Human-Computer Interaction. We present here our \gui prototype. The back-end of the prototype is a \python Web server. It is integrated with the observation execution system of \cta, which is based on the Alma Common Software (\acs). The back-end incorporates a \redis database, which facilitates synchronization of \gui panels. \redis is also used to buffer information collected from various software components and databases. The front-end of the prototype is based on Web technology. Communication between Web server and clients is performed using \webSock, where graphics are generated with the \djs \jvs library.
\end{abstract}

\section{Introduction}
  %
  The Cherenkov Telescope Array (\cta)~\cite{2011ExA....32..193A} is
  the next generation observatory
  for very high-energy gamma-rays (\gamrays). \cta
  will be sensitive to photon energies from $20$~\gev, up to a few hundred~\tev.
  As they impact the atmosphere, \gamrays induce particle showers.
  A part of these cascades is Cherenkov radiation,
  which is emitted as charged particles travel faster than the speed of light in the atmosphere.
  The Cherenkov radiation can be detected by imaging atmospheric
  Cherenkov telescopes (\iacts)~\cite{Hillas201319}.
  Using multiple telescopes together, the particle showers can stereoscopically be reconstructed.
  This allows one to derive the properties of the initial \gamray that initiated the shower.

  \cta will include three types of primary instruments. These
  correspond to three telescope sizes,
  large-, mid- and small-size, respectively called \lsts, \msts, and \ssts.
  The different telescope types will be sensitive to different \gamray energy ranges,
  where smaller mirror areas correspond to higher energies.
  \cta telescope arrays will be deployed in two sites in the southern and northern hemispheres,
  respectively consisting of~$\Sim100$ and~$\Sim20$ telescopes.
  This represents a substantial increase in the number of instruments
  compared to existing \iact
  experiments (\hess~\cite{2006A&A...457..899A}, \veritas~\cite{Holder:2006gi},
  and \magic~\cite{Albert:2007xh}),
  which incorporate between two to five telescopes.

  The focus of this paper is the graphical user
  interface (\gui) for the operator of a \cta site.
  %
  As mentioned above, \cta will include a large number of telescopes. In addition,
  several variations of \sst and \mst designs will be
  used. This further increases the diversity of hardware which needs to be controlled by an operator.
  The complexity of \cta makes for an interesting challenge in designing an effective
  user interface.
  In the following, the development process of the operator \gui is detailed.
  We describe the way in which the \gui is foreseen to be integrated
  with other \cta systems. In addition, we showcase a few of the features of the
  current prototype implementation\footnote{
    Media resources illustrating panels from the prototype
    are available at \protect\url{https://www-zeuthen.desy.de/~sadeh/}\;.
  }.

\section{Design and development process}
  %
  Nominally, on-site observing operations with \cta will be automated. The purpose of
  the operator \gui can therefore be summarised as follows:
  %
  \begin{Enumerate}{}
  \item
     start and end observations;
  \item
     override the automated scheduled operations in order to perform a specific
     task, or for safety reasons (could require manual control over instruments);
  \item
     monitor the state of the array during operations, including
     hardware status and simple scientific metrics;
  \item
     identify, diagnose, and if possible resolve, problems with specific systems or processes.
  \end{Enumerate}

  The development process of the \gui is inspired by the model used by
  the Atacama Large Millimeter/submillimeter Array (\alma)~\cite{2009IEEEP..97.1463W}.
  \alma is an astronomical interferometer of radio telescopes in the Atacama desert of northern Chile.
  It is similar in complexity to \cta,
  comprising~$66$ radio antennas of three general types.
  One of the main lessons learned from \alma, is that it is important to take into account
  advances in Human-Computer Interaction (\hci) when designing
  an interface~\cite{pietriga:hal-00735792}.

  The design process is thus driven by participatory workshops.
  These bring together several groups of people: experts from the field of \hci;
  experienced telescope operators; control software experts; and astroparticle physicists.
  The purpose of the workshops is
  to refine the requirements on the \gui;
  to better understand what the \gui should enable users to accomplish;
  and to work out how best to implement these ideas, without causing unnecessary cognitive load.
  The outcome of the first two workshops, including a set of initial requirements,
  is described in detail in~\cite{Sadeh:2016kje}.

\section{Incorporating the \gui within Central-Control}
  %
  \subsection{A Prototype for the \gui}
    %
    The current prototype for the operator \gui is based on HTML5 Web technologies.
    The back-end of the prototype is a \python server, based on the
    \pyramid\cite{pyramid} framework.
    %
    The front-end is displayed in a Web browser. The design
    is implemented using \polymer\cite{polymer},
    a \webComp application programming interface,
    developed by \google.
    Data are displayed using an open-source \jvs library,
    called \djs\cite{d3js}.
    Asynchronous communication between the back-end and the
    front-end of the GUI is performed using \webSock.

  \subsection{The Observation Execution System}
    %
    The operator \gui is part of the Observation Execution System (\oes) of \cta.
    The purpose of \oes is
    to monitor and control telescopes and auxiliary devices;
    to schedule and execute observations and calibrations;
    and to time-stamp, read-out, filter, and store the collected data.
    \oes is comprised of several sub-systems in addition to the \gui.
    These are:
    \tit{Manager and Central Control};
    \tit{Short-Term and \too} (target of opportunity) \tit{Scheduler};
    \tit{Data Handling};
    \tit{Configuration};
    \tit{Reporting and Diagnosis};
    and \tit{Monitoring}.

    At the high-level, \oes is implemented upon the
    Alma Common Software (\acs) framework.
    \acs is a distributed middleware framework, based on a container-component model.
    It supports \python, \java, and \cpp, and
    constitutes a thoroughly tested platform for distributed data acquisition
    and instrumentation control.

  \subsection{Interfaces Between the \gui and \oes}
    %
    Interfaces between the \gui and each of the sub-systems of \oes are already envisioned.
    For example:
    \hlt{-~Manager and Central Control:}
    The \gui will monitor the status of the
    \tit{observing blocks} which are currently being
    executed. The latter are logical units, which define a specific observing or calibration task. They
    include information, such as
    the time-span in which to carry out the task, the required resources (\eg telescopes), etc.
    \hlt{-~Short-term and \too Scheduler:}
    The \gui will have access to the plan for future observations. It will also have
    the functionality to modify this plan on the short-term, by \eg creating or cancelling observing blocks.
    \hlt{-~Monitoring:}
    The \gui will display all monitoring data which are \eg deemed necessary for
    discovering and diagnosing potential hardware problems.

    The main users of the \gui will be a single or several operators, who
    will occupy the \cta control-room.
    However, additional users are also foreseen, such as off-site engineers.
    The potentially large number of users increases the load on \oes.
    For example,
    one may consider the case of a detailed monitoring view for a telescope. For
    such a panel, hundreds of data-points are displayed,
    each updated with a $O(1~\txt{Hz})$ rate.
    A simple architecture may entail a communication channel being opened between
    the \gui and the telescope, for each one of the requested properties.
    Such an approach has an obvious downside;
    it would result in a large demand on resources,
    whenever multiple users access the same panel at the same time.

    In order to avoid such problems, a part of the infrastructure
    of the \gui will be a dedicated database.
    The primary purpose of the latter is to act as a buffer, and reduce the traffic between
    the \gui and other \oes components.
    For the current prototype, we chose \redis for this purpose.
    \redis is an in-memory data structure store, which can achieve high write and read speeds.
    \redis has several advantages over other database solutions. Primarily,
    it is optimised for quickly
    updating and querying large numbers of non-complex data elements, such as strings.

    %
    %
    \begin{figure}[t]
      \centering
      \includegraphics[trim=0mm 70mm 142mm 0mm,clip,width=0.4\textwidth]{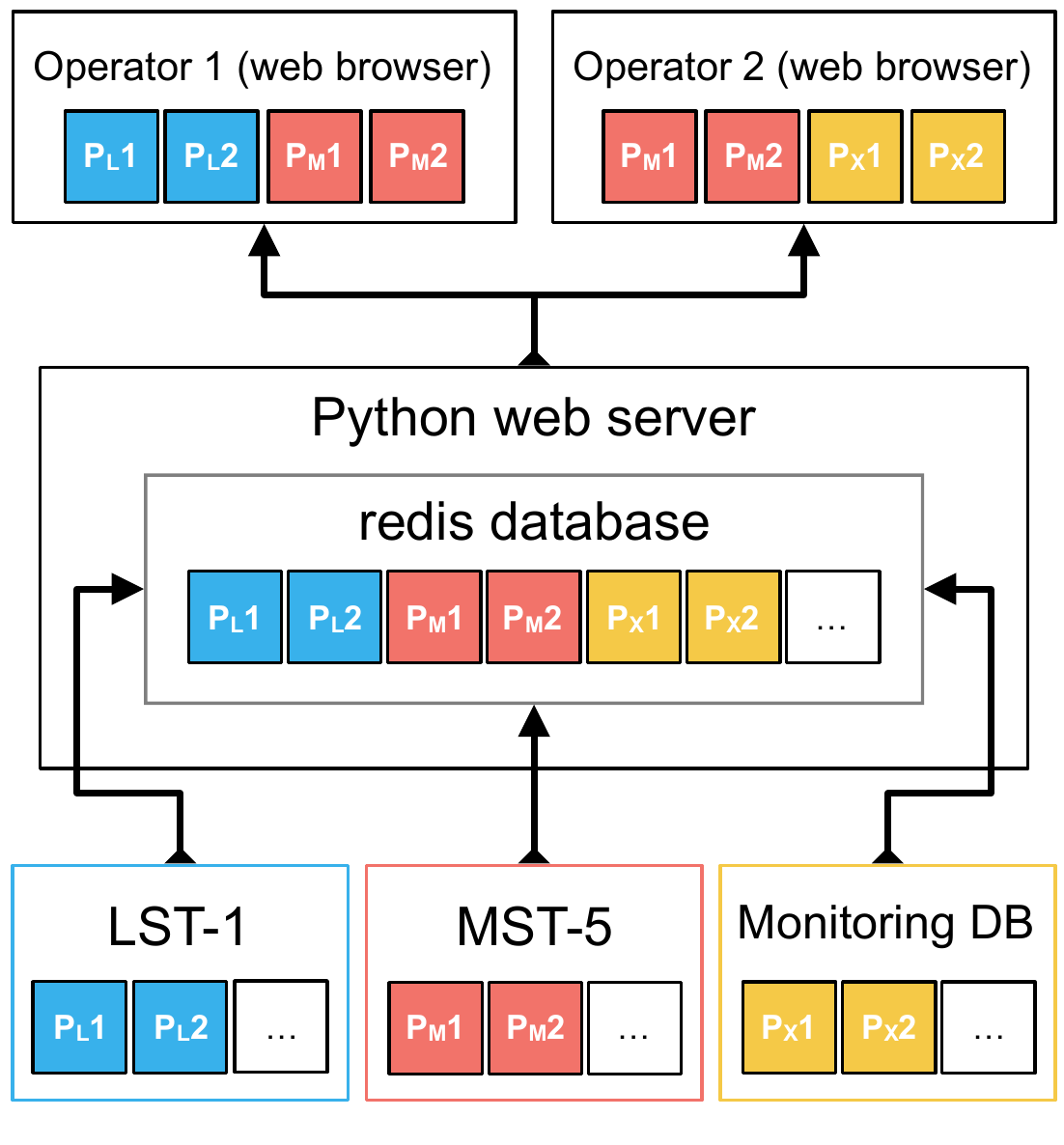}
      \caption{Schematic
        illustration of the mechanism for exposing monitoring data to users of the \gui.
        In this example, two users (Operators~1 and~2) request data from a combination of sources. The
        data transmitted to the users are first stored in a \redis database, which is integrated with the \python Web server.
        \redis is filled by directly acceding lower-level elements of the system. In this example, these are
        two telescopes (\lst-1 and \mst-5), and the monitoring database, an \oes component.}
      \label{FIGguiArch}
    \end{figure}
    \FigRef{FIGguiArch} shows a schematic illustration of the functionality of the
    \redis database.
    In this example, two users request monitoring information.
    The users may ask for the same, or for different data. In any case,
    communication between the users and \oes is mediated
    through \redis, avoiding direct links to other part of the system.
    In turn, \redis is filled by accessing lower-level components.
    In this example, the data partly originate from a direct link to telescopes,
    and partly from the monitoring database (another component of \oes).

    Considering the scope of the data stored in the database,
    two general approaches for resource management are possible.
    One option is for \redis to continuously be filled by
    all available properties which could potentially be requested by any panel of the \gui.
    Alternatively, properties may be monitored on-demand.
    That is, specific data will only be stored and updated in \redis,
    if they are required by at least a single user at a given time.
    The first option is more simple, and therefore less prone to errors.
    On the other hand, the second path has the advantage of
    requiring significantly less resources.
    Both configurations are currently being explored.

  \section{Front-end design features of the \gui prototype}
    
    \subsection{Example Telescope Monitoring Panel}

      \begin{figure*}[p]
        \centering
          \begin{minipage}[c]{1\textwidth}
            \begin{minipage}[c]{0.5\textwidth}
              \centering\subfloat[]{\label{FIGarrayZoomer1}\includegraphics[trim=1mm -20mm 1.5mm -5mm,clip,width=.98\textwidth]{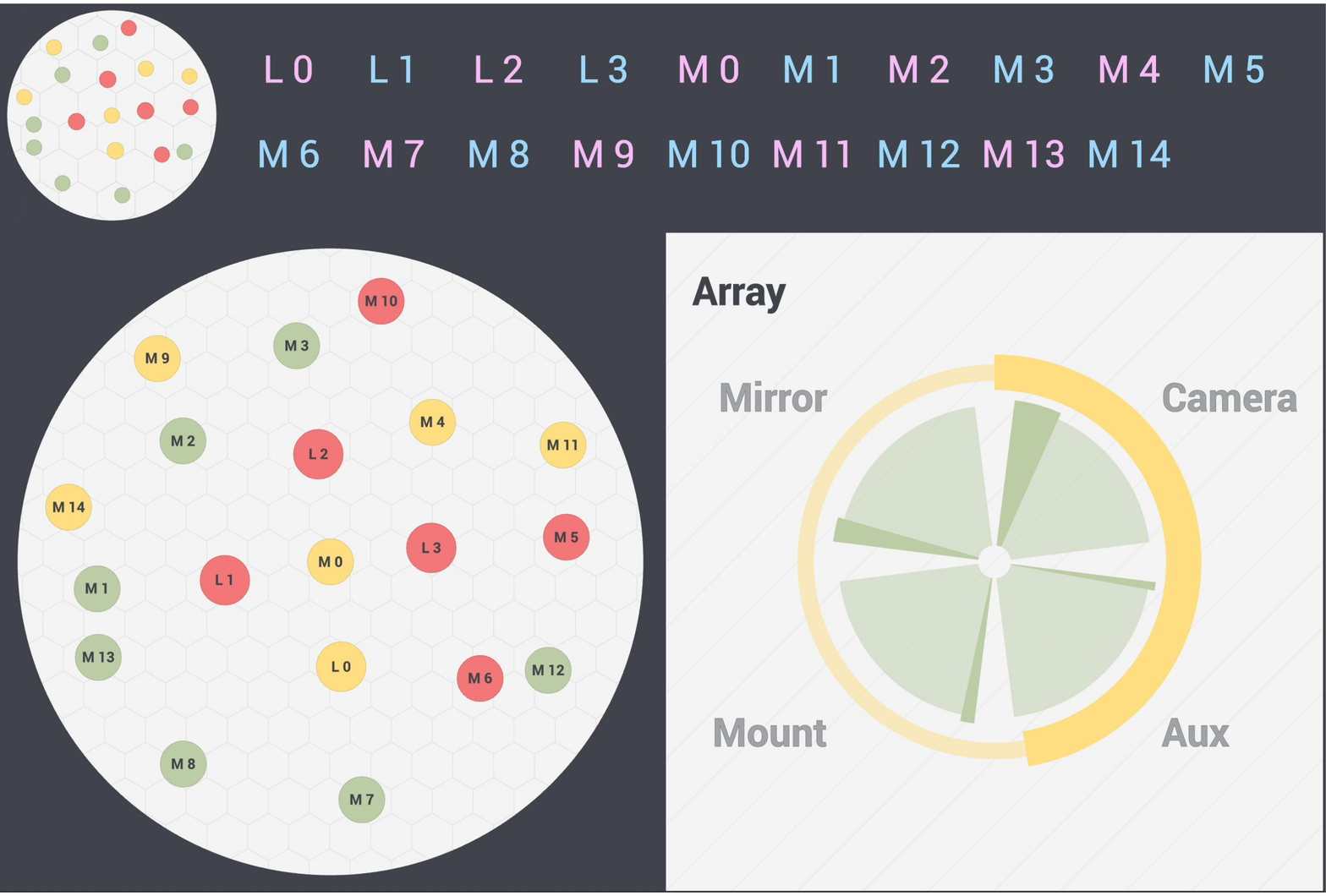}}
            \end{minipage}\hfill
            \begin{minipage}[c]{0.5\textwidth}
              \centering\subfloat[]{\label{FIGarrayZoomer2}\includegraphics[trim=1mm -20mm 1.5mm -5mm,clip,width=.98\textwidth]{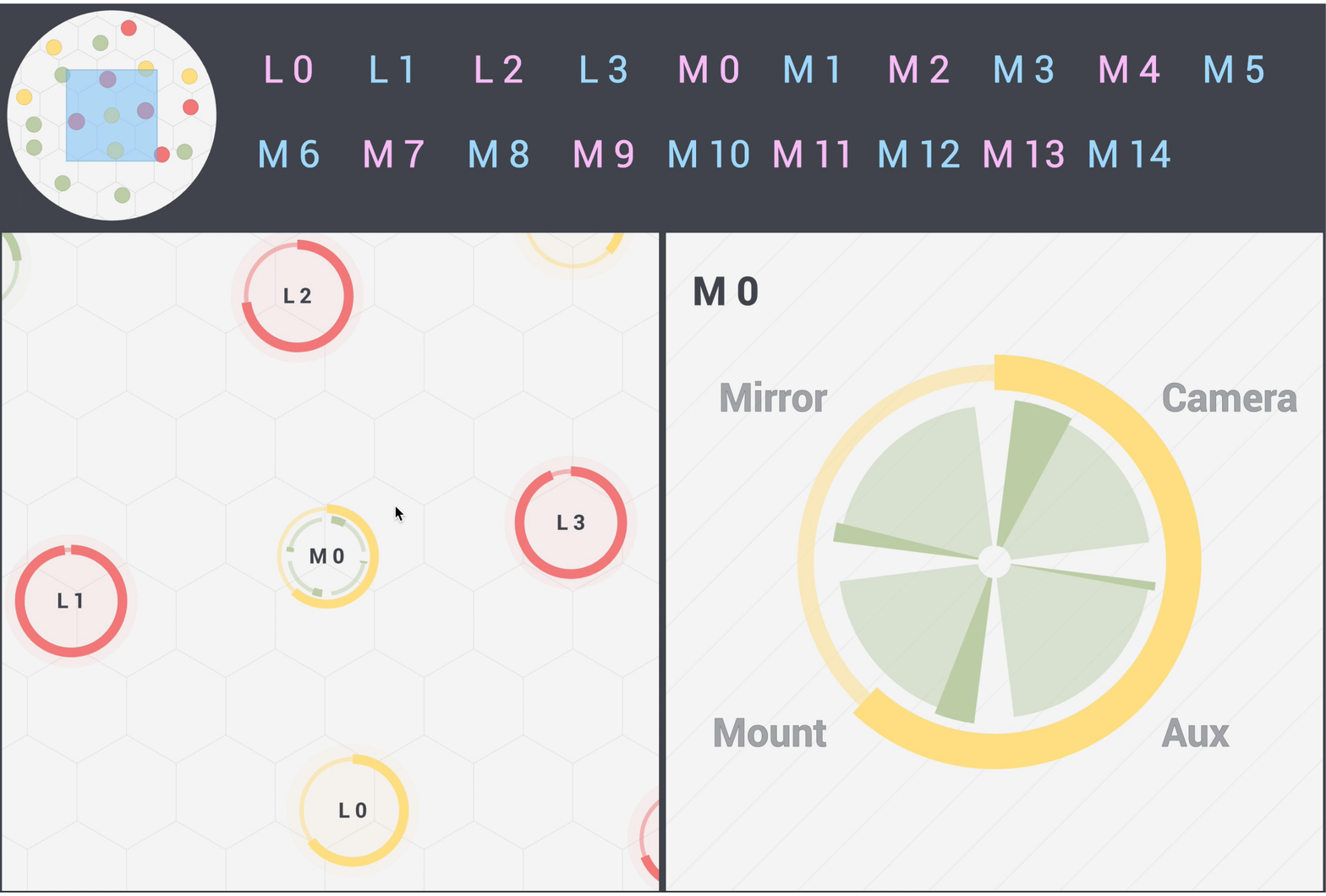}}
            \end{minipage}\hfill
          \end{minipage}\hfill
          \begin{minipage}[c]{1\textwidth}
            \begin{minipage}[c]{0.5\textwidth}
              \centering\subfloat[]{\label{FIGarrayZoomer3}\includegraphics[trim=1mm -20mm 1.5mm -5mm,clip,width=.98\textwidth]{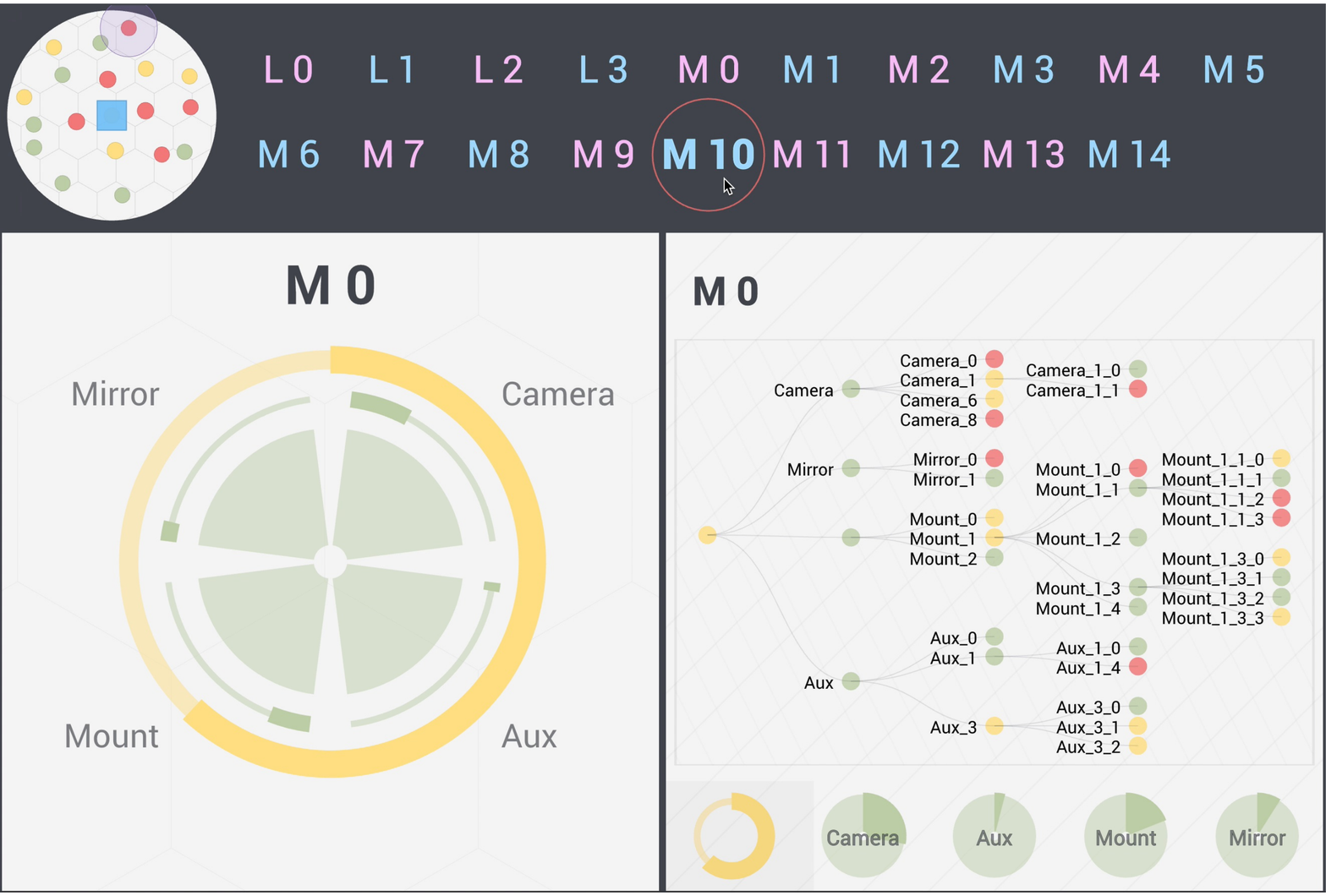}}
            \end{minipage}\hfill
            \begin{minipage}[c]{0.5\textwidth}
              \centering\subfloat[]{\label{FIGarrayZoomer7}\includegraphics[trim=1mm -20mm 0mm -5mm,clip,width=.98\textwidth]{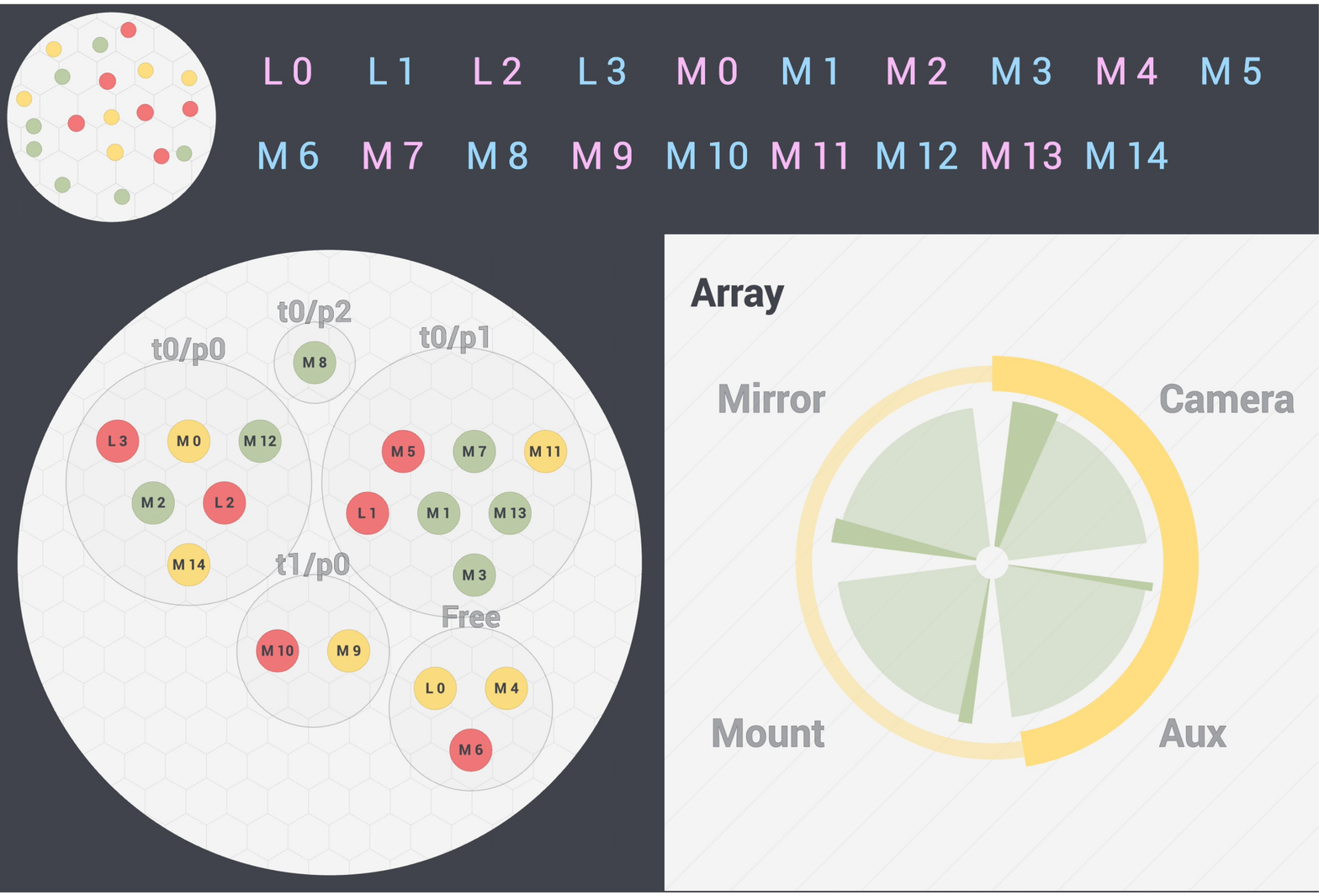}}
            \end{minipage}\hfill
          \end{minipage}\hfill
          \vspace{8pt}
          \begin{minipage}[c]{1\textwidth}
            \begin{minipage}[c]{0.333\textwidth}
              \centering\subfloat[]{\label{FIGarrayZoomer4}\includegraphics[trim=1mm -30mm 1.5mm 57mm,clip,width=.98\textwidth]{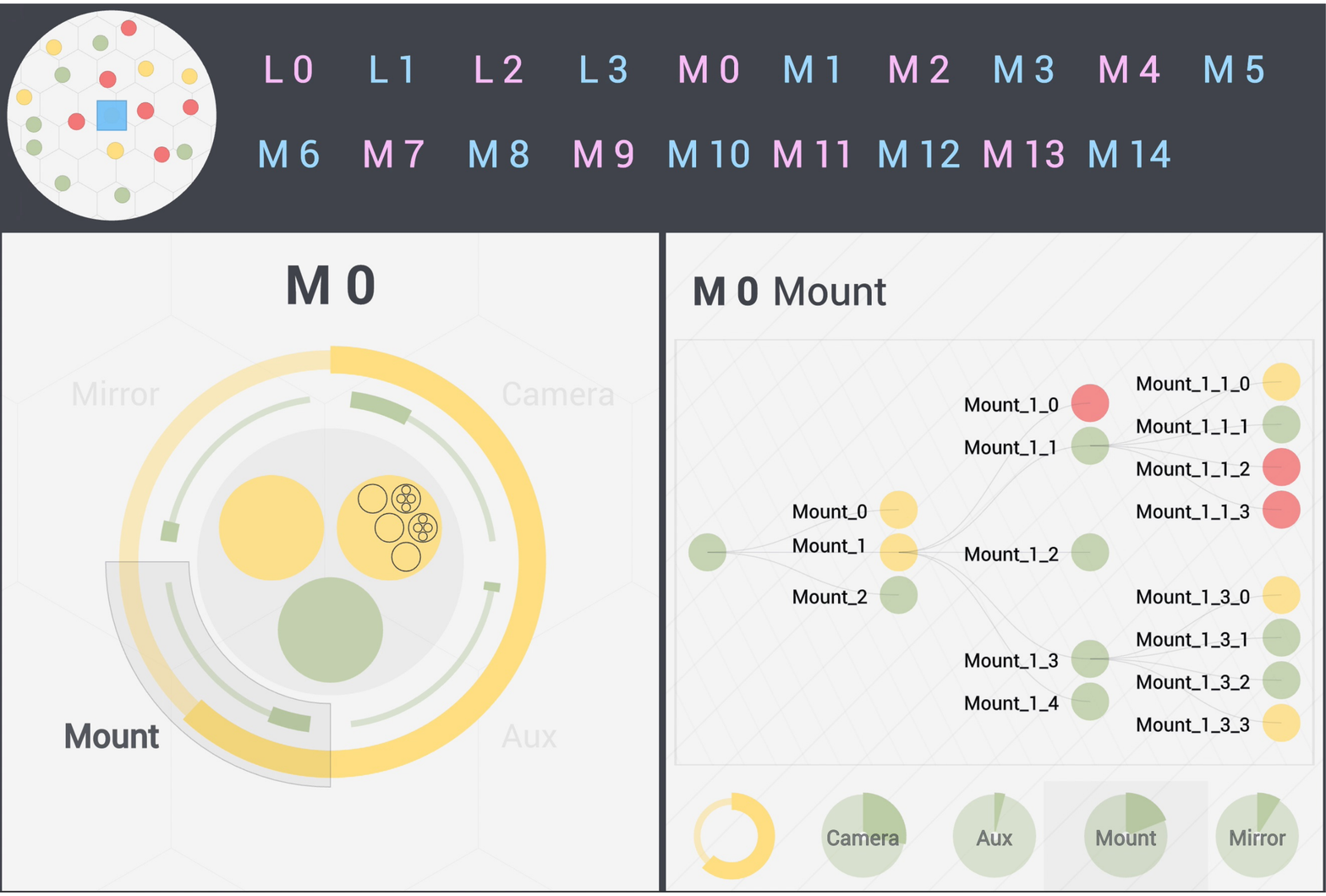}}
            \end{minipage}\hfill
            \begin{minipage}[c]{0.333\textwidth}
              \centering\subfloat[]{\label{FIGarrayZoomer5}\includegraphics[trim=1mm -30mm 1.5mm 57mm,clip,width=.98\textwidth]{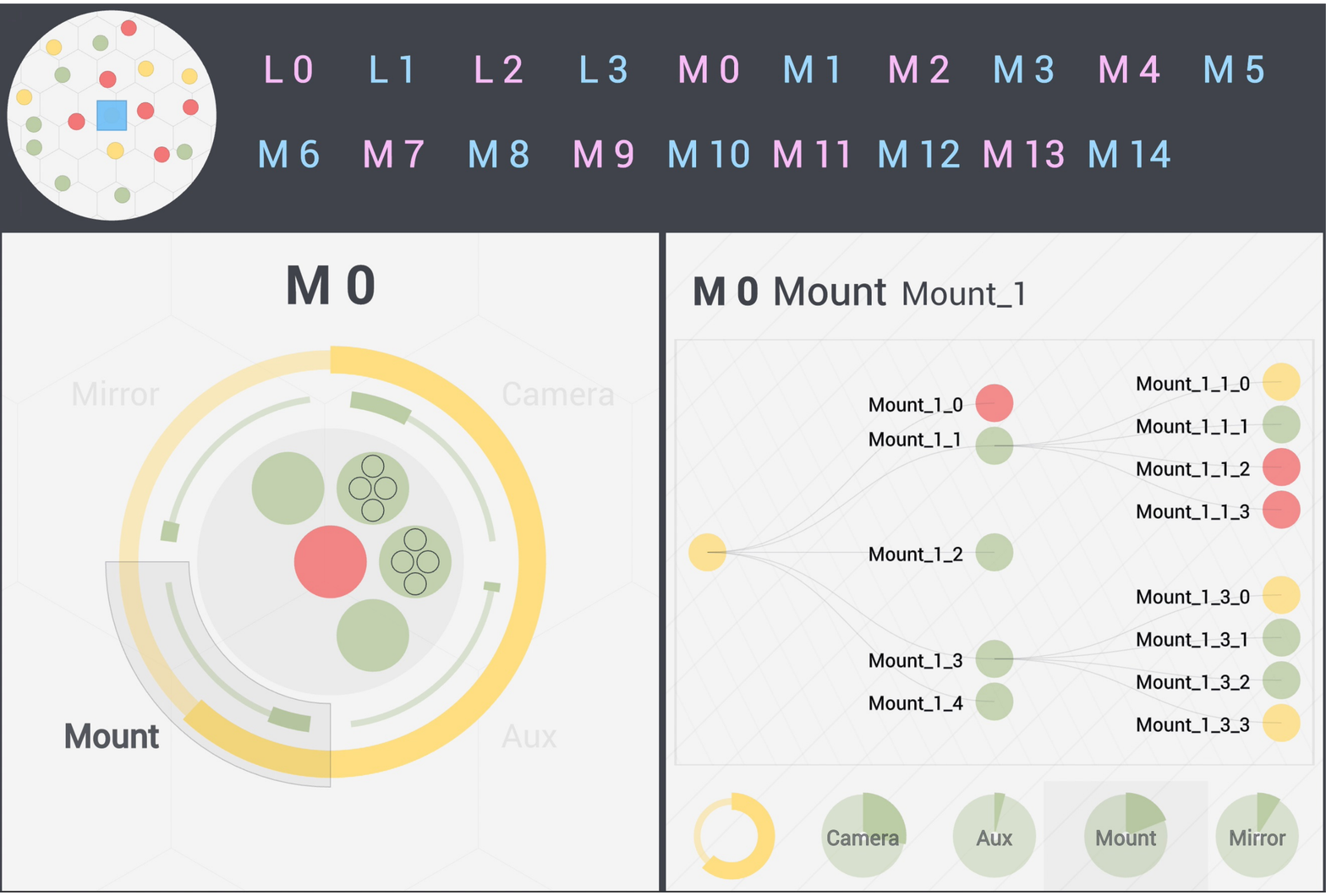}}
            \end{minipage}\hfill
            \begin{minipage}[c]{0.333\textwidth}
              \vspace{2.8pt}
              \centering\subfloat[]{\label{FIGarrayZoomer6}\includegraphics[trim=1mm -35mm 1.5mm 57mm,clip,width=.98\textwidth]{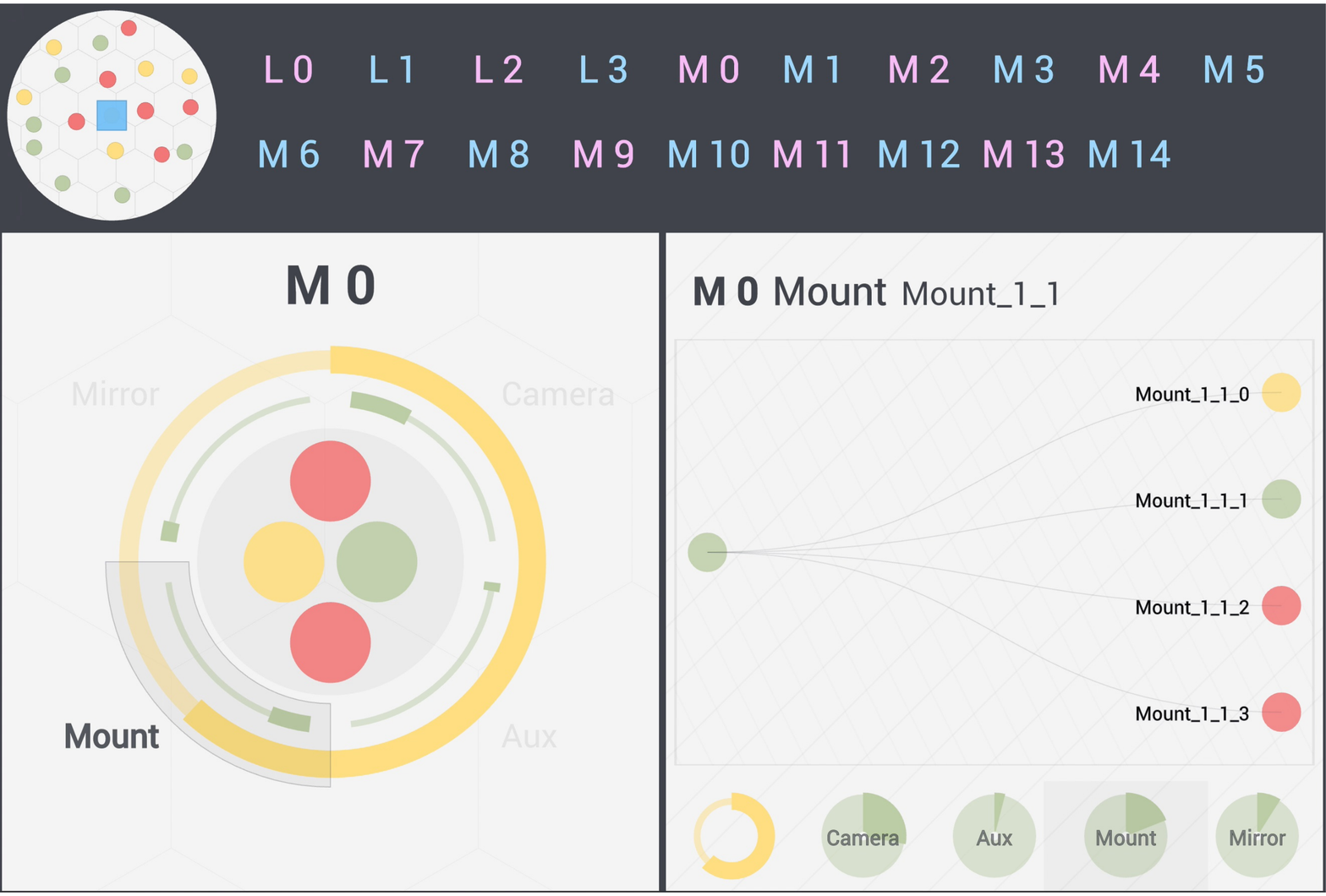}}
            \end{minipage}\hfill
          \end{minipage}\hfill
          \vspace{10pt}
          \caption{Several view-points and configurations of a telescope monitoring panel, as discussed in the text.}
          \label{FIGarrayZoomer}
      \end{figure*}

      Several panels have already been implemented as part of the prototype of the operator \gui.
      As an example, a telescope monitoring panel is presented in \figRef{FIGarrayZoomer}, using
      the array layout of the northern \cta site. The view is composed of several
      sub-panels, as described in the following.

      Several view-points of the same panel are shown in the figure.
      \FigSubRef{FIGarrayZoomer}{FIGarrayZoomer1} illustrates the initial view. The top left corner contains
      a \tit{mini-map}, which is a fixed perspective of the entire array. Each circle represents a single
      telescope, where the different units are arranged in a pseudo-geographic layout.
      The colours represent global status indicators for each telescope,
      green (normal operations), yellow (some problems), and red (device failure).
      %
      The wide dark area on the top (from centre to right corner) is a \tit{chess-board}. The latter allows
      for quick selection of specific telescopes, based on their ID tag. The chess-board complements
      the functionality of the mini-map, which enables selection based on physical location.

      The bottom right sub-panel in \figSubRef{FIGarrayZoomer}{FIGarrayZoomer1} shows a general
      status indicator for the entire array. In addition to a global metric, indicated by the outer yellow ring, four
      quadrants are shown, tagged as: \tit{Mirror}, \tit{Camera}, \tit{Aux} (auxiliary), and \tit{Mount}. These
      represent a generalized division of a telescope into four sub-systems, which have their
      own common functionality.
      The status of a given element is represented by two redundant indicators:
      the colours (green to red);
      and the relative shaded area in the outer ring, or in each of the inner quadrants.

      The bottom left sub-panel in \figSubRef{FIGarrayZoomer}{FIGarrayZoomer1} shows a larger view of the
      pseudo-geographic layout of the array. Unlike the static mini-map, the larger view
      includes zooming functionality. The effect of zooming can be understood from
      \FigSubRef{FIGarrayZoomer}{FIGarrayZoomer2}.
      In this case, telescope \mst-0 (indicated by ${\txt{M\;0}}$) is selected and zoomed upon.
      The zoom action is also reflected in the mini-map as a shaded blue rectangle, which allows for
      the pseudo-geographic context to be kept.
      In addition, the bottom-right view reflects the selection action; the global array metrics have been
      replaced with the status indicators of the selected telescope, ${\txt{M\;0}}$.

      The panel incorporates \tit{semantic zooming}~\cite{Perlin1993}.
      The latter is a technique that enables representing information at
      different levels of detail for a given visual element.
      This behaviour complements standard geometric zooming,
      where only the size of elements changes with the zoom-factor.
      As the view is zoomed-in on a specific area, the graphical representation is
      changed for each one of the enclosed telescopes, increasing the level of detail.

      As the zoom-factor is further increased by the user,
      the display is updated again, as indicated in \figSubRef{FIGarrayZoomer}{FIGarrayZoomer3}.
      The bottom left sub-panel shows the different metrics for each sub-system of ${\txt{M\;0}}$. The bottom right
      sub-panel shows a \tit{hierarchical tree diagram} of the different components associated with each sub-system.
      For the current prototype, these components are not yet assigned to specific hardware elements.
      Place-holder names, such as \tit{\ttt{Mount\under0}}, are used instead.
      \FigSubRef{FIGarrayZoomer}{FIGarrayZoomer3} also shows a secondary functionality
      of the chess-board. As the user hovers over a
      given telescope, the latter is highlighted by a thin circle within the mini-map. This allows
      for a quick association between a specific telescope and its position within the map, even when the larger
      display is zoomed-in.

      The two main sub-panels (bottom left and right) are synchronized using brushing and linking,
      following the principles of
      \tit{coordinated multiple views}~\cite{WangBaldonado:2000:GUM:345513.345271,North00}.
      A user-action in one is reflected in the other.
      The display allows a user to focus on a specific sub-system,
      by clicking on one of the quadrants from the bottom left display, or
      on one of the circles in the tree on the right.
      This is shown in \figSubRef{FIGarrayZoomer}{FIGarrayZoomer4} for the \tit{Mount}.
      It is possible to further traverse the hierarchy of the tree, and focus on lower levels
      of components. Three such levels are shown in \figsSubRef{FIGarrayZoomer}{FIGarrayZoomer5}-\Subref{FIGarrayZoomer6}.
      In addition to collapsing the tree into lower \tit{branches}, the hierarchy is also
      represented by the enclosed circles inside the sub-panel on the left.
      This \tit{circle-packing} display~\cite{Wang:2006:VLH:1124772.1124851}
      allows for quick navigation, to higher or to lower branches of the tree.

      The large array display on the bottom left can also be modified, such that telescopes are not
      mapped to a pseudo-geographic layout. \FigSubRef{FIGarrayZoomer}{FIGarrayZoomer7} illustrates
      one such arrangement, where the
      mini-map, chess-board and the detailed element view on the bottom right all remain unchanged.
      In this example, telescopes are grouped into \tit{sub-arrays}, represented as an \tit{enclosure diagram}.
      A sub-array is a collection of telescopes which perform a single task, corresponding to an observing block
      (\eg observing an astronomical object).
      Regardless of the change in layout, the zoom-properties of the different
      visual elements in the display remain the same, preserving the basic functionality of the panel.

    \subsection{Example Observing Block Panel}
      %
      %
      \begin{figure*}[tp]
        \centering
          \begin{minipage}[c]{0.85\textwidth}
        	\centering
              \includegraphics[trim=0mm -10mm 0mm 0mm,clip,width=0.95\textwidth]{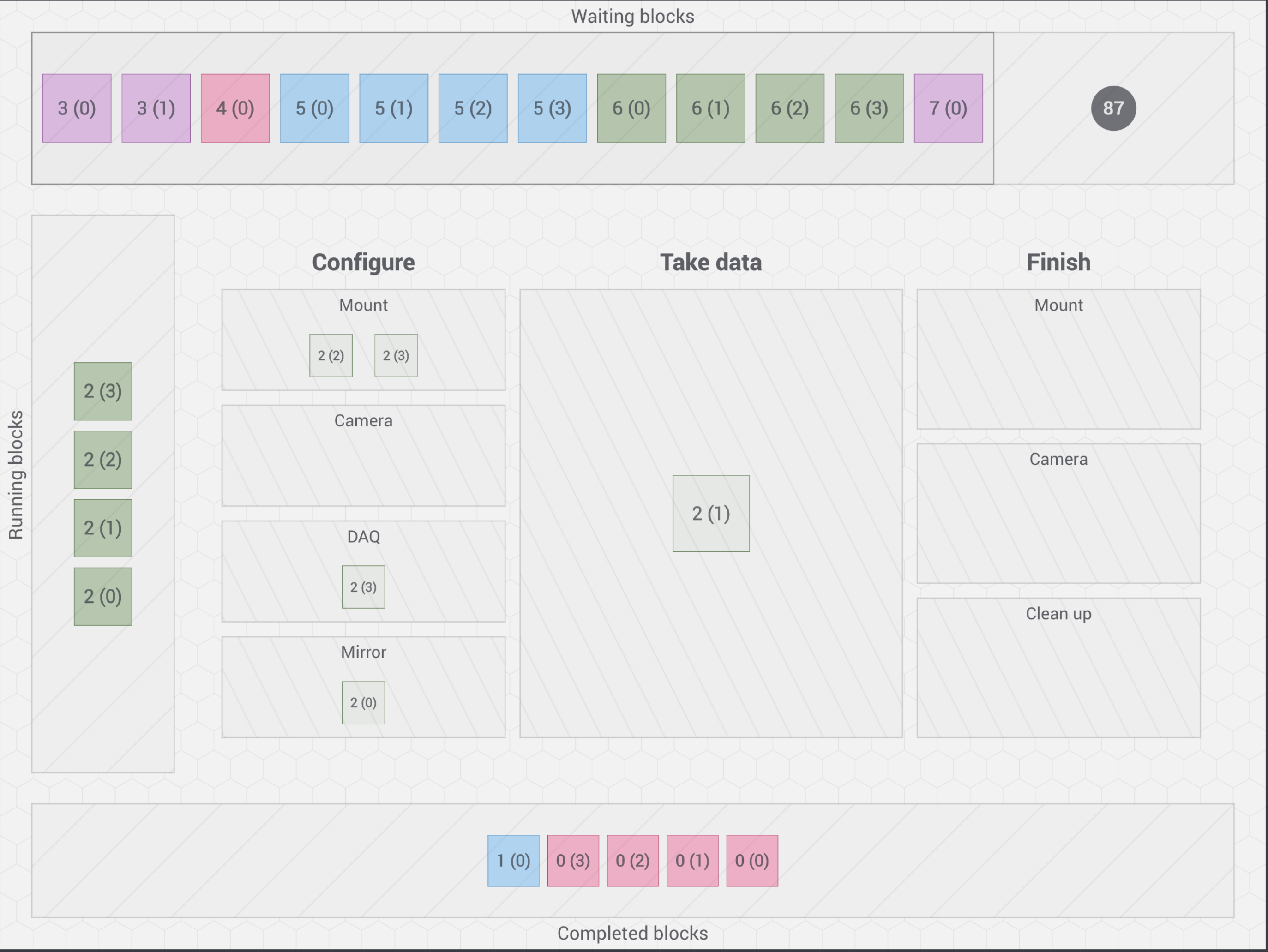}
          \end{minipage}
          \vspace*{-\baselineskip}
          \caption{An observing block monitoring panel.}
          \label{FIGobsBlocks}
      \end{figure*}
      Another example of a panel from the prototype is presented in \figRef{FIGobsBlocks}.
      The purpose of this panel is to display
      the current execution status of different observing blocks.
      The different blocks are represented as squares in the panel.
      Each square is positioned within a rectangle, that indicates the execution status.
      The latter includes three categories:
      \tit{waiting blocks} (top);
      \tit{running blocks} (middle-left); and
      \tit{completed blocks} (bottom).
      In addition, the middle section of the panel contains \tit{phase} groups, which
      provide a detailed breakdown of the status of running blocks. These include
      a \tit{configure} phase of the various telescope sub-systems;
      a \tit{take data} phase, to indicate that \eg observations are being performed;
      and a \tit{finish} phase, where the different sub-systems go through
      any procedures which are \eg needed as an observation ends.

      As a user clicks on a given observing block within the view, detailed information
      regarding the block is displayed in another panel. The latter
      is currently being designed, and so is not shown here.
      An interface panel is also being developed, which will allow a user
      to create new observing blocks with customized parameters, and to cancel
      waiting or running blocks.

    \subsection{Synchronization of Panels}
      %
      An important aspect of the design of the \gui is panel-synchronization.
      The purpose of synchronization is to enable robust interaction
      of a user with the \gui~\cite{WangBaldonado:2000:GUM:345513.345271,North00}.
      A simple example of synchronization, is the coordinated behaviour of the two
      main sub-panels shown in \figRef{FIGarrayZoomer}.
      However, the full benefits of this technique are found in interconnecting
      separate panels, where different aspects of the system can share the same context.

      To illustrate this, one may consider for example the
      connection between the telescope monitoring and the observing block panels.
      The two are synchronized, such that an event is transmitted from the
      telescope monitoring panel, when the user zooms-in on a particular
      telescope. This event is received by the observing block panel; it has
      the effect of highlighting the corresponding observing block,
      to which the selected telescope is assigned.

      Work is currently under way to also extend the scope
      of the telescope monitoring panel. The objective is to synchronize it
      with a new view, which will contain monitoring plots.
      \begin{figure*}[thp]
        \centering
          \begin{minipage}[c]{1\textwidth}
            \includegraphics[trim=0mm -5mm 0mm 10mm,clip,width=.98\textwidth]{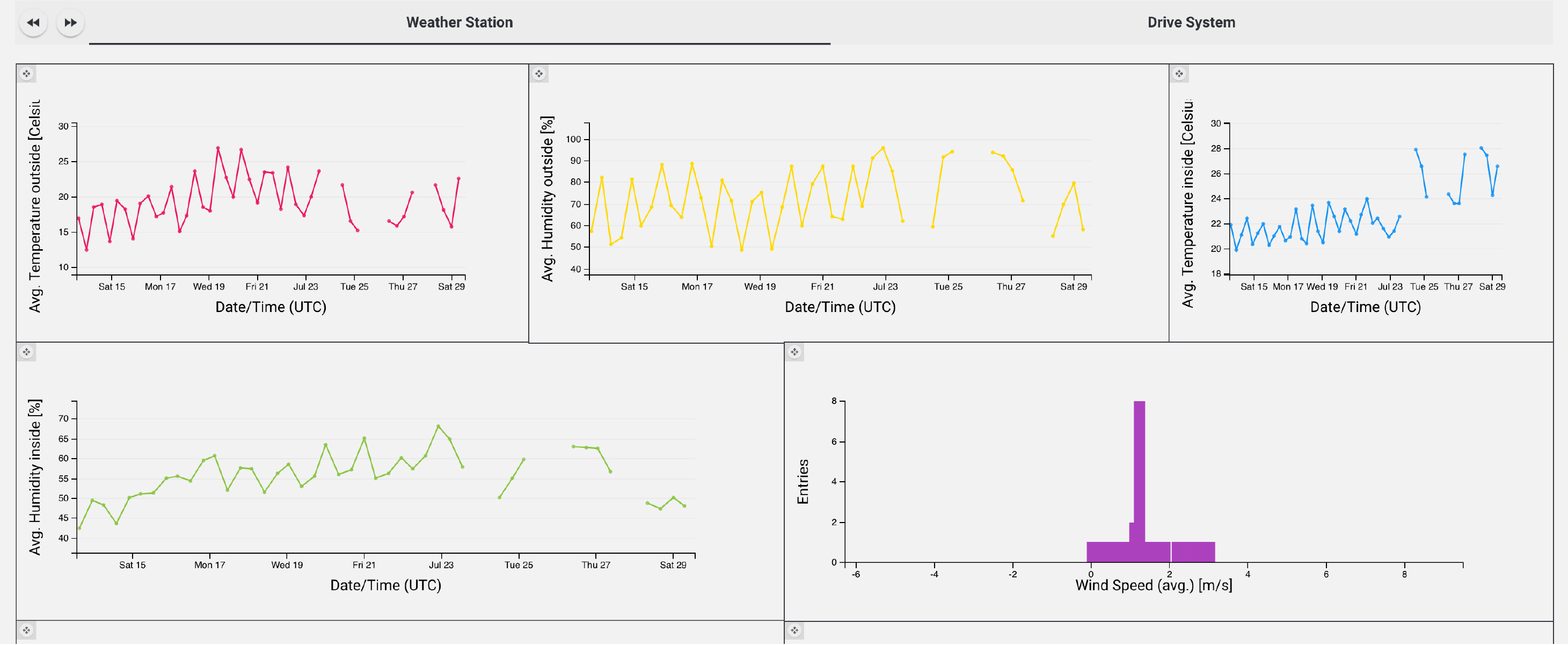}
          \end{minipage}\hfill
          \caption{A partial example of a monitoring plot panel.}
          \label{FIGmonitoringPlots}
      \end{figure*}
      A partial example of such a panel is shown in \figRef{FIGmonitoringPlots}.
      In this case, the plots show time-series and a histogram, derived from data taken
      with a weather monitoring station.

      For the purpose of a synchronised view with the panel shown in \figRef{FIGarrayZoomer},
      the data would correspond to each of the monitoring-points within the tree diagram.
      By clicking on a given element from the tree, such as \ttt{Mount\under0}, the respective
      distributions of the value of \ttt{Mount\under0} over time would appear.
      The plots would be arranged in such a way, that the hierarchy of monitored
      data elements will be visually emphasized. This will \eg facilitate tracing
      of problems across a range of inter-connected hardware components.

  \section{Summary}
    %
    The design of an effective \gui
    for the operator of \cta is an interesting and challenging task, due to the
    complexity of the observatory.
    This paper presents the status of the design process of the
    current prototype implementation of the \gui.
    Two aspect of the prototype are discussed, the back-end, and the front-end.

    The back-end of the \gui is a \python Web server. Some of the interfaces
    of the server with the Observation Execution System of \cta are discussed.
    In particular, a description is given of
    the choice of using a \redis database as a buffering layer
    between the \gui and other \oes components.

    The front-end of the \gui is a displayed in a Web browser.
    Two of the existing panels from the prototype are showcased,
    a detailed telescope monitoring view, and an observing block monitor.
    Some of the specific design details of these panels are illustrated,
    such as semantic zooming. In addition, a description is given of
    a general feature, panel synchronization, which allows
    different views to be interconnected.

    Future work will include implementing additional interfaces between the \gui
    and the rest of \oes.
    Concerning the front-end, existing \gui panels will
    be extended with \eg new synchronization behaviour.
    New views will also be designed, in order to cover the full required
    functionality of the \gui, as briefly summarised above.

  \section{Acknowledgement}
    %
    We would like to thank C.~Appert, E.~Barrios, A.~Bulgarelli, A.~Cabrera, F.~Dazzi,
    G.~Fasola, V.~Fioretti, L.~Font, M.~Fuessling, S.~Gabici,
    M.~Garczarczyk, K.~Kosack, R.~Macias, R.~White and the members of the
    \cta \oes team, for their helpful comments and insights.

    We gratefully acknowledge financial support from the agencies and organizations listed here:\\
    \url{http://www.cta-observatory.org/consortium_acknowledgments}.



\end{document}